\documentclass[preprint,12pt,authoryear]{elsarticle}

\usepackage{amssymb}
\usepackage{graphicx}
\usepackage{braket}
\usepackage{amsmath}
\usepackage{hyperref}
\usepackage{float}
\usepackage{amsthm}
\usepackage{longtable}
\usepackage{booktabs}
\usepackage{CJKutf8}
\usepackage[strings]{underscore}
\usepackage{braket}
\usepackage{enumitem}
\usepackage{diagbox}
\usepackage{tikz}
\usepackage{bm} 
\usepackage[outline]{contour} 
\usepackage{subcaption} 
\contourlength{0.25pt}
\usetikzlibrary{shapes} 
\usetikzlibrary{shadows.blur}
\colorlet{myred}{red!85!black!65}
\colorlet{mycyan}{blue!40!cyan!85!black!55}
\colorlet{myblue}{blue!70!cyan!85!black!55}
\colorlet{mydarkblue}{myblue!90!black!90}
\colorlet{mydarkerblue}{myblue!80!black!90}
\colorlet{mydarkestblue}{myblue!70!black!90}
\colorlet{mylightblue}{blue!50!cyan!80!black!45}
\colorlet{mypink}{magenta!95!red!100!black!72}
\colorlet{mypurple}{blue!50!red!90!black!70}
\colorlet{mydarkpurple}{blue!50!red!70!black!70}
\colorlet{mygreen}{green!60!black!75}
\colorlet{mylightgreen}{green!80!red!90!black!50}
\colorlet{myorange}{orange!95!black!70}
\colorlet{mybrown}{brown!80!orange!95!black!80}
\colorlet{myyellow}{yellow!80!orange!95!black!80}
\colorlet{mylightyellow}{yellow!80!orange!95!black!45}

\def\R{1} 
\tikzstyle{pin}=[very thin,line cap=round]
\tikzset{
  slice/.style={fill=#1,draw=#1!80!black,line join=round,line width=0.4,
                blur shadow={shadow blur steps=20,shadow xshift=0.5,
                             shadow yshift=-1,shadow opacity=40}},
  slice/.default=myblue,
  pics/piechart/.style n args={2}{ 
    code={
      \foreach \frac/\name/\col [
        count=\i,
        remember=\angb as \anga (initially #1), 
        evaluate={\angm=\anga-\frac*1.8;  
                  \angb=\anga-\frac*3.6;  
                  \exp=\R*max(0.02,0.2*(1-\frac/100)^8); 
                  \r=\exp+\R*max(0.5,(0.8-\frac/100));} 
      ] in {#2}{ 
        \coordinate (P\i) at (\angm:\exp+\R);
        \draw[slice=\col] 
          (\angm:\exp) --++ (\anga:\R) arc(\anga:\angb:\R) -- cycle;
        \node[white,scale=0.9]
          at (\angm:\r) {\contour{\col!75!black}{\bm{$\frac\mathbf{\%}$}}};
        \node[\col!70!black,anchor=180+\angm,inner sep=4]
          at (P\i) {\bf\bm{$\name$}};
      }
    }
  },
  pics/piechart2/.style n args={2}{ 
    code={
      \foreach \frac/\name/\col [
        remember=\angb as \anga (initially #1), 
        evaluate={\angm=\anga+\frac*1.8;  
                  \angb=\anga+\frac*3.6;  
                  \exp=\R*max(0.02,0.14*(1-\frac/100)^3); 
                  \r=\exp+\R*max(0.5,(1-\frac/100)^2);} 
      ] in {#2}{ 
        \message{^^J frac=\frac: anga=\anga -> angb=\angb}
        \draw[slice=\col] 
          (\angm:\exp) --++ (\anga:\R) arc(\anga:\angb:\R) -- cycle;
        \ifdim \frac pt > 10pt 
          \node[white,align=center]
            at (\angm:\r) {\boldlabel{$\name$}{\frac}};
        \else 
          \node[\col!5!black,align=center,anchor=180+\angm,inner sep=3]
            at (\angm:\exp+\R) {\boldlabel{$\name$}{\frac}};
        \fi
      }
    }
  },
}

\newcommand\boldlabel[2]{\bf\bm{#1}\\[-1]\small\bm{$#2\mathbf{\%}$}}

\begin{document}
\begin{CJK}{UTF8}{gbsn}
\begin{frontmatter}



\title{Exploring the Potential of Quantum Approximate Optimization Algorithm in Tackling the Perfect Domination Problem}


\author[a,b]{Haoqian Pan\corref{cor1}\fnref{cm}}
\author[a]{Changhong Lu}
\author[c]{Yuqing Zheng}
\author[d]{Chunxing Yan}

\affiliation[a]{organization={School of Mathematical Sciences,  Key Laboratory of MEA(Ministry of Education) \& Shanghai Key Laboratory of PMMP, East China Normal University}, 
            city={Shanghai},
            postcode={200241},
            country={China}}

\affiliation[b]{
  organization={China Communications Information \& Technology Group Co., Ltd.},
  city={Beijing},
  postcode={101399},
  country={China}
 }  

\affiliation[c]{organization={School of Astronomy and Space Science, University of Science and Technology of China}, 
            city={Hefei},
            postcode={230026},
            country={China}}

\affiliation[d]{organization={School of Finance and Mathematics, Huainan Normal University}, 
            city={Huainan},
            postcode={232038},
            country={China}}
 
\cortext[cor1]{Corresponding author.}
\fntext[cm]{52215500040@stu.ecnu.edu.cn}

\begin{abstract}

Perfect Domination Problem (PDP), a canonical challenge in combinatorial optimization, finds critical applications in real-world systems such as error-correcting codes, wireless communication networks, and social networks. Decades of research have firmly established its NP-completeness across numerous graph classes. Motivated by rapid advances in quantum computing, significant effort has recently been directed toward quantum algorithms for NP-complete problems, most notably the Quantum Approximate Optimization Algorithm (QAOA). Nonetheless, the applicability and efficacy of quantum approaches to the PDP remain entirely unexplored. This paper initiates the first systematic investigation of the PDP via QAOA. We evaluate solution quality on three benchmark instances of 6, 7, and 8 vertices using 15–18 qubits on a quantum simulator, examining more than 400 distinct parameter configurations. Experimental results confirm the algorithm’s effectiveness and expose discernible trends in parameter selection. These outcomes substantiate QAOA’s viability for the PDP and mark a seminal step toward situating this classical problem within the quantum-computing paradigm.
  
\end{abstract}


\end{frontmatter}


\section{Introduction}\label{sec:Introduction}

For a graph \(G=(V,E)\), a dominating set (DS) is a subset \(D\subseteq V\) such that every vertex \(v\in V\setminus D\) is adjacent to at least one vertex in \(D\). The classical Domination Problem (DP) asks for a DS of minimum cardinality. A more restrictive criterion yields the perfect dominating set (PDS): here, every vertex \(v\in V\setminus D\) must be adjacent to exactly one vertex in \(D\). The Perfect Domination Problem (PDP) seeks a PDS of smallest possible size. Structurally, the PDP is a direct refinement of the DP; nevertheless, the two problems have distinct origins. The DP emerged from facility-location frameworks \citep{chang2013algorithmic}, whereas the PDP arose as a graph theoretic generalisation of error-correcting codes in communication theory \citep{biggs1973perfect}. Beyond these foundational contexts, the PDP finds wide-ranging applications in parallel computing architectures \citep{RN482}, wireless communication networks \citep{RN483}, and social network analysis \citep{RN485}.

From an algorithmic perspective, PDP has been proven NP-complete for many graph classes \citep{RN486, RN487}. Over the years, numerous algorithms have been developed to address the PDP, often tailored to specific graph types or specialized problem variations, such as the weighted version. Significant studies have explored the PDP in various graph categories, including rectangular grid graphs \citep{RN488}, distance-hereditary graphs \citep{RN493}, circular-arc graphs \citep{RN490}, and knight’s graphs \citep{RN492}. Additionally, the weighted PDP has been investigated in tree graphs \citep{RN487}, chordal graphs, and split graphs \citep{RN491}, while weighted independent PDP solutions have been studied for cocomparability graphs \citep{RN489}. Despite some algorithms achieving polynomial or even linear time complexity for restricted graph types, solving the PDP remains computationally challenging in its general form due to its NP-complete nature.

In recent years, the rapid advancement of quantum computing \citep{RN421, RN495, RN496, RN497} has spurred the development of numerous quantum algorithms aimed at addressing combinatorial optimization problems. Prominent among these algorithms are Quantum Annealing \citep{RN422}, the Quantum Approximate Optimization Algorithm (QAOA) \citep{RN436}, and the Variational Quantum Eigensolver \citep{RN428}, among others. As one of the most influential quantum algorithms of the past decade, QAOA has been applied to a wide array of combinatorial optimization problems, including max-cut \citep{RN436}, the traveling salesman problem \citep{RN453}, the DP \citep{RN415,zhang2024quantum}, max-flow \citep{RN420}, the minimum vertex cover problem \citep{RN334}, the boolean satisfiability problem\citep{RN455}, among others. Nevertheless, as a pivotal problem bridging communications and combinatorics, PDP has, to date, not been approached by any quantum algorithm; investigations into quantum algorithms for this problem remain entirely absent.

Against this backdrop, we investigate the deployment of the QAOA for the resolution of PDP. Employing 14–18 qubits within a quantum simulator, we successfully solved PDP on graphs of varying scales. The principal contributions of this work are as follows. (1) We present the first quantum‐algorithmic treatment of PDP; experimental evidence confirms that the QAOA-based framework is viable for the problem. (2) We conducted exhaustive tests across extensive parameter landscapes, with selected instances spanning up to 420 distinct parameter configurations. The resulting data elucidate the response patterns and performance tendencies of QAOA when applied to PDP, thereby furnishing critical insights and concrete guidance for subsequent research.

The structure of the paper is as follows. In Section \ref{sec:problemmodeling}, we present a 0-1 integer programming model for the PDP based on its definition and outline the steps to transform it into a Hamiltonian. In Section \ref{sec:QAOA}, we introduce the fundamental concept of QAOA. In Section \ref{sec:Experiment}, we employ a quantum simulator to solve the PDP using QAOA under various parameter combinations and provide a comprehensive analysis of the experimental results. Finally, in Section \ref{sec:conclusion}, we conclude the paper with a summary of the key findings.

\section{Problem modeling of PDP} \label{sec:problemmodeling}
The premise of using QAOA is to first map the objective function of a combinatorial optimization problem to the Hamiltonian of a quantum system. In this process, the Quadratic Unconstrained Binary Optimization (QUBO) model often acts as a bridge. The standard approach involves transforming the original problem into, or directly modeling it as, a QUBO model, which is then converted into a Hamiltonian through variable substitution. Following this methodology, our approach sequentially transforms the PDP into a 0-1 integer programming model, a QUBO model, and finally a Hamiltonian. We begin by presenting the 0-1 integer programming model for the PDP.
\begin{alignat}{2}
  \min_{\{X_{i}\}} \quad & \sum\limits_{i=1}^{|V|} X_{i} \label{eq:pdptarget},\\
  \mbox{s.t.}\quad
  &\sum\limits_{j \in N[i]} X_{j} \ge 1 \quad \forall i \in V  \label{eq:pdpc1},\\
  &X_{i} \in \{0,1\}  \quad \forall i \in V, \\
  &\sum\limits_{ij \in E} [X_{i} \cdot (1 - X_{j}) + X_{j} \cdot (1 - X_{i}) ]  = |V| - \sum\limits_{i}^{|V|} X_{i} \label{eq:pdpc2}.
\end{alignat}
The decision variable \( X_{i} \) is defined such that \( X_{i} = 1 \) if \( i \in D \), and \( X_{i} = 0 \) otherwise. Accordingly, the objective function in Eq. \ref{eq:pdptarget} represents the total number of vertices in \( D \). Constraint \ref{eq:pdpc1} ensures that for each vertex \( i \), either \( i \) is in \( D \) or it has at least one neighbor in \( D \). This constraint guarantees that \( D \) forms a DS. For constraint \ref{eq:pdpc2}, the left-hand side calculates the number of edges in \( E \) where one endpoint belongs to \( V \setminus D \) and the other belongs to \( D \). The right-hand side represents the number of vertices in \( V \setminus D \). Since \( D \) is a DS, every vertex in \( V \setminus D \) must have at least one neighbor in \( D \). Thus, the number of edges connecting \( V \setminus D \) to \( D \) must be at least \( |V| - \sum\limits_{i}^{|V|} X_{i} \). When this inequality holds as an equality, it signifies that each vertex in \( V \setminus D \) has exactly one neighbor in \( D \), satisfying the perfect domination condition. Therefore, constraint \ref{eq:pdpc2} is the defining constraint of the PDP.

After formulating the 0-1 integer programming model for the PDP, the next step is to transform it into a QUBO model. The standard representation of the QUBO model is provided in Eq. \ref{eq:qubo}, where the matrix \( Q \), commonly referred to as the QUBO matrix.
\begin{equation}
  minimize/maximize \quad y = x^{t}Qx \label{eq:qubo}.
\end{equation}
Given that \( X_{*} \in \{0, 1\} \), it follows that \( X_{*} = X_{*}^{2} \). As a result, the objective function of the 0-1 integer programming model for the PDP inherently satisfies the requirements of the QUBO model. The next step involves transforming the constraints into quadratic penalty terms. The general form of constraint \ref{eq:pdpc1} is expressed as follows:
\begin{equation}
  X_{1} + X_{2} + \dots + X_{n} \geq 1, \quad n = |N[i]|, \quad \forall i \in V \label{eq:ctnormal}. 
\end{equation}
According to \cite{RN416}, for \( n = 1 \) or \( n = 2 \), the constraint can be transformed into \( P \cdot (X_{1} - 1)^{2} \) and \( P \cdot (1 - X_{1} - X_{2} + X_{1} \cdot X_{2}) \), respectively, where \( P \) is the penalty coefficient. The value of \( P \) typically requires adjustment based on the specific characteristics of the problem. As suggested in \cite{RN416}, setting \( P \) to 0.75 to 1.5 times the value of the original objective function serves as a reasonable starting point. For \( n \geq 3 \), slack variables must be introduced to convert the inequality in Eq. \ref{eq:ctnormal} into an equality constraint, as shown in Eq. \ref{eq:scpdom}.
\begin{equation}
  X_{1} + X_{2} + \dots + X_{n} - S - 1 = 0 \label{eq:scpdom}.
\end{equation}
It is evident that the range of \( S \) encompasses all integers within the interval \([0, n-1]\). To represent \( S \), we introduce additional 0-1 variables, following the formulation provided in Eq. \ref{eq:sc} \citep{RN420, RN501}. This approach ensures that \( S \) can take any integer value within the specified range \([0, n-1]\).
\begin{equation}
  S = \sum\limits_{i=1}^{bl_{n-1}-1} X_{i}^{\prime}\cdot 2^{i-1} + (n - 1 - \sum\limits_{i=1}^{bl_{n-1}-1}2^{i-1}) \cdot X_{bl_{n-1}}^{\prime}. \label{eq:sc}
\end{equation}
In Eq. \ref{eq:sc}, \( X_{*}^{\prime} \in \{0,1\} \) represents the newly introduced binary variables, while \( bl_{n-1} \) denotes the length of the binary representation of \( n-1 \). This approach, as utilized by \cite{RN420}, has been applied to represent flow values in the maximum flow problem, with its correctness formally established in \cite{RN501}. Building on these works, we transform Eq. \ref{eq:scpdom} into quadratic penalties, as expressed in Eq. \ref{eq:qp}.
\begin{equation}
  P \cdot (X_{1} + X_{2} + \dots + X_{n} - [\sum\limits_{i=1}^{bl_{n-1}-1} X_{i}^{\prime} \cdot 2^{i-1} + (n - 1 - \sum\limits_{i=1}^{bl_{n-1}-1}2^{i-1}) \cdot X_{bl_{n-1}}^{\prime}] - 1) ^ {2} \label{eq:qp}
\end{equation}
It is easy to see that when \( D \) is a DS, we have
\begin{equation}
  \sum\limits_{ij \in E} [X_{i} \cdot (1 - X_{j}) + X_{j} \cdot (1 - X_{i}) ]  \geq |V| - \sum\limits_{i}^{|V|} X_{i} \label{eq:pdomqp2}.
\end{equation}
Therefore, we can convert it without using the square form into
\begin{equation}
  P \cdot (\sum\limits_{ij \in E} [X_{i} \cdot (1 - X_{j}) + X_{j} \cdot (1 - X_{i}) ]  - |V| + \sum\limits_{i}^{|V|} X_{i}).
\end{equation}
Since constraint \ref{eq:pdpc2} builds upon the premise of constraint \ref{eq:pdpc1}, we assign distinct symbols to represent their respective penalty coefficients, \( P_{2} \) and \( P_{1} \), where \( P_{2} \leq P_{1} \). By transforming the two types of constraints in the PDP into quadratic penalty terms, we ultimately derive the QUBO model for the PDP (Eq. \ref{eq:pdomqubo}). Next, to proceed with converting the QUBO model into a Hamiltonian, we replace \( X \) and \( X^{\prime} \) with binary variables \( s \), which take values in \( \{-1, 1\} \). The conversion process is outlined in Eq. \ref{eq:sx}. 
\begin{equation}
  \begin{split} 
  &\min_{\{X,X^{\prime}\}} \quad  \sum\limits_{i=1}^{|V|} X_{i}  \\
  &+ \sum\limits_{i \in V, |N[i]| \geq 3} P_{1} \cdot [\sum\limits_{j \in N[i]}X_{j}  - (\sum\limits_{i=1}^{bl_{|N[i]|-1}-1} X_{i}^{\prime} \cdot 2^{i-1} + (|N[i]| - 1 - \sum\limits_{i=1}^{bl_{|N[i]|-1}-1}2^{i-1}) \cdot X_{bl_{|N[i]|-1}}^{\prime}) - 1]^{2} \\
  & + \sum\limits_{i \in V, |N[i]| = 2, N[i] = \{j,k\}} P_{1} \cdot (1- X_{j} - X_{k} + X_{j} \cdot X_{k}) \\
  & + \sum\limits_{i \in V, |N[i]| = 1, N[i] = \{j\}} P_{1} \cdot (X_{j} - 1) ^ {2}\\
  & + P_{2} \cdot (\sum\limits_{ij \in E} [X_{i} \cdot (1 - X_{j}) + X_{j} \cdot (1 - X_{i}) ]  - |V| + \sum\limits_{i}^{|V|} X_{i}).
  \label{eq:pdomqubo}
  \end{split}
\end{equation}
\begin{equation}
  X_{i} = \frac{s_{i} + 1}{2} \label{eq:sx}.
\end{equation}
After the substitution, we have Eq. \ref{eq:pdomqubos}. Finally, by replacing \( s \) and \( s^{\prime} \) with the Pauli-Z operator \( \sigma^{z} \), we can ultimately obtain the Hamiltonian (Eq. \ref{eq:hpdom}). At this stage, we have completed the process of transforming the PDP from a 0-1 integer programming model into a Hamiltonian. In the subsequent sections, we will outline the basic workflow of QAOA.

\begin{equation}
  \begin{split} 
  &\min_{\{s,s^{\prime}\}} \quad  \sum\limits_{i=1}^{|V|} \frac{s_{i} + 1}{2}  \\
  &+ \sum\limits_{i \in V, |N[i]| \geq 3} P_{1} \cdot [\sum\limits_{j \in N[i]}\frac{s_{j} + 1}{2}\\  
  &- (\sum\limits_{i=1}^{bl_{|N[i]|-1}-1} \frac{s_{i}^{\prime} + 1}{2} \cdot 2^{i-1} + (|N[i]| - 1 - \sum\limits_{i=1}^{bl_{|N[i]|-1}-1}2^{i-1}) \cdot \frac{s_{bl_{|N[i]|-1}}^{\prime} + 1}{2}) - 1]^{2} \\
  & + \sum\limits_{i \in V, |N[i]| = 2, N[i] = \{j,k\}} P_{1} \cdot (1- \frac{s_{j} + 1}{2} - \frac{s_{k} + 1}{2} + \frac{s_{j} + 1}{2} \cdot \frac{s_{k} + 1}{2}) \\
  & + \sum\limits_{i \in V, |N[i]| = 1, N[i] = \{j\}} P_{1} \cdot (\frac{s_{j} + 1}{2} - 1) ^ {2} \\
  & + P_{2} \cdot (\sum\limits_{ij \in E} [\frac{s_{i} + 1}{2} \cdot (1 - \frac{s_{j} + 1}{2}) + \frac{s_{j} + 1}{2} \cdot (1 - \frac{s_{i} + 1}{2}) ]  - |V| + \sum\limits_{i}^{|V|} \frac{s_{i} + 1}{2})
  \label{eq:pdomqubos}.
  \end{split}
\end{equation}
\begin{equation}
  \begin{split} 
  &\hat{H}_{c} = \sum\limits_{i=1}^{|V|} \frac{\hat{\sigma}_{i}^{z} + 1}{2}   \\
  &+ \sum\limits_{i \in V, |N[i]| \geq 3} P_{1} \cdot [\sum\limits_{j \in N[i]}\frac{\hat{\sigma}_{j}^{z} + 1}{2}\\  
  &- (\sum\limits_{i=1}^{bl_{|N[i]|-1}-1} \frac{(\hat{\sigma}_{i}^{z})^{\prime} + 1}{2} \cdot 2^{i-1} + (|N[i]| - 1 - \sum\limits_{i=1}^{bl_{|N[i]|-1}-1}2^{i-1}) \cdot \frac{(\hat{\sigma}_{bl_{|N[i]|-1}}^{z})^{\prime} + 1}{2}) - 1]^{2} \\
  & + \sum\limits_{i \in V, |N[i]| = 2, N[i] = \{j,k\}} P_{1} \cdot (1- \frac{\hat{\sigma}_{j}^{z}+ 1}{2} - \frac{\hat{\sigma}_{k}^{z} + 1}{2} + \frac{\hat{\sigma}_{j}^{z} + 1}{2} \cdot \frac{\hat{\sigma}_{k}^{z} + 1}{2}) \\
  & + \sum\limits_{i \in V, |N[i]| = 1, N[i] = \{j\}} P_{1} \cdot (\frac{\hat{\sigma}_{j}^{z} + 1}{2} - 1) ^ {2} \\
  & + P_{2} \cdot (\sum\limits_{ij \in E} [\frac{\hat{\sigma}_{i}^{z} + 1}{2} \cdot (1 - \frac{\hat{\sigma}_{j}^{z} + 1}{2}) + \frac{\hat{\sigma}_{j}^{z} + 1}{2} \cdot (1 - \frac{\hat{\sigma}_{i}^{z} + 1}{2}) ]  - |V| + \sum\limits_{i}^{|V|} \frac{\hat{\sigma}_{i}^{z} + 1}{2}).
  \label{eq:hpdom}
  \end{split}
\end{equation}

\section{QAOA}\label{sec:QAOA}
To begin, the spin operator and the Pauli operator are related as follows:
\begin{equation}
  \hat{s} = \frac{\hbar}{2} \hat{\sigma}.
\end{equation}
For the \( z \)-component of \( \hat{s} \), it is naturally related to \( \hat{\sigma}^{z} \). In the Pauli representation,
\begin{equation}
  \hat{\sigma}^{z} = \begin{pmatrix}
    1 & 0 \\
    0  & -1\\
    \end{pmatrix}.
\end{equation}
It has two eigenstates, \( \ket{0} = \begin{pmatrix} 1 \\ 0 \end{pmatrix} \) and \( \ket{1} = \begin{pmatrix} 0 \\ 1 \end{pmatrix} \), which correspond to two distinct spin directions. Initially, we assume that each qubit in a quantum system with \( n \) qubits is in the \( \ket{0} \) state, such that the system's state is represented as \( \ket{\underbrace{00\ldots 0}_{n}} \). Subsequently, QAOA applies the Hadamard gate (Eq. \ref{eq:hgate}) to prepare this state into an equal superposition of all basis states, as described in Eq. \ref{eq:is}. Here, the bit string \( z = z_{1} z_{2} z_{3} \dots z_{n} \), where \( z_{i} \in \{0, 1\} \).

\begin{equation}
  \hat{H} = 2^{-\frac{1}{2}} ([\ket{0} + \ket{1}] \bra{0} + [\ket{0} -\ket{1}] \bra{1}) \label{eq:hgate}.
\end{equation}

\begin{equation}
  \ket{s} = \underbrace{\hat{H} \otimes \hat{H} \cdots \otimes \hat{H}}_{n} \ket{\underbrace{00\dots 0}_{n}} = \frac{1}{\sqrt{2^{n}} } \cdot \sum\limits_{z} \ket{z} \label{eq:is}.
\end{equation}
Next, QAOA applies two types of rotation operators, \( U(\hat{C}, \gamma) \) and \( U(\hat{B}, \beta) \) (Eq. \ref{eq:ug}, Eq. \ref{eq:ub}), to the initial state \( \ket{s} \), repeating the process \( q \) times. Here, \( \hat{C} = \hat{H}_c \), \( \hat{B} = \sum\limits_{j=1}^{n} \hat{\sigma}_{j}^{x} \), with \( \gamma \in [0, 2\pi] \) and \( \beta \in [0, \pi] \). Let \( \gamma_{q} \) and \( \beta_{q} \) denote the angles used in the \( q \)-th layer. After \( q \) iterations, the final state \( \ket{\gamma, \beta} \) is obtained, as shown in Eq. \ref{eq:gammabeta}.
\begin{align}
  U(\hat{C},\gamma) &= e^{-i\gamma \hat{H}_{c}} \label{eq:ug}.\\
  U(\hat{B},\beta) &=  e^{-i\beta \hat{B}} \label{eq:ub}.
\end{align}
\begin{equation}
  \ket{\gamma,\beta} = U(\hat{B},\beta_{q})U(\hat{C},\gamma_{q}) \cdots U(\hat{B},\beta_{1})U(\hat{C},\gamma_{1}) \ket{s} \label{eq:gammabeta}.
\end{equation}
With fixed values of \( \gamma \) and \( \beta \), by repeatedly executing the quantum circuit depicted in Eq. \ref{eq:gammabeta} and measuring the final state, the expectation value of \( \hat{H}_{c} \), denoted as \( F_{q}(\gamma, \beta) \), can be obtained.
\begin{equation}
  F_{q}(\gamma,\beta) = \bra{\gamma,\beta} \hat{H}_{c} \ket{\gamma,\beta}.
\end{equation}
Since the PDP is a minimization problem, the values of \( \gamma_{*} \) and \( \beta_{*} \) at each layer must be adjusted to minimize \( F_{q}(\gamma, \beta) \). The essence of QAOA lies in applying two types of operators in a manner that increases the probability of the system collapsing into basis states corresponding to the smallest values of \( H_c \). As a hybrid algorithm, QAOA leverages classical optimization methods, such as COBYLA or Nelder-Mead, to tune the angles at each layer. The optimization process terminates when either the maximum number of iterations is reached or the function tolerance falls below a predefined threshold. Once the optimal values of \( \gamma_{*} \) and \( \beta_{*} \) are determined, the quantum circuit is updated accordingly, and multiple final samples are generated. The bit string \( z_{*} \) with the highest probability from these samples is then output. In \( z_{*} \), the spin direction of each qubit corresponds to the 0-1 values of the decision variables in the original optimization problem, allowing for the direct extraction of the final PDS. According to theoretical results \citep{RN436}, increasing the number of layers \( q \) brings \( F_{q}(\gamma, \beta) \) closer to the optimal value. However, deeper layers lead to more complex quantum circuits, which present significant challenges when implementing QAOA on both real quantum computers and local quantum simulators. 


In this chapter, we introduced the basic workflow of QAOA. In the subsequent chapter, we will conduct experiments to evaluate the performance of QAOA in solving the PDP.
\section{Experiment}\label{sec:Experiment}

\subsection*{Environment}

This experiment utilized Qiskit for quantum simulation. The CPU used in this experiment was an AMD R9 7950X3D, paired with 48GB of memory. Qiskit was employed to construct the quantum circuit, simulate the backend, and perform sampling. We adopted the angle initialization method proposed in \cite{RN458} and used COBYLA as the optimization algorithm, with a tolerance of \( 10^{-8} \). All code for this work was implemented in Python. The primary parameters for this experiment included the layer number \( q \), penalty coefficients \( P_{1} \) and \( P_{2} \), and the maximum number of iterations. 

Throughout the experimental phase, we employed three graphs comprising 6, 7, and 8 vertices, respectively, corresponding to 14, 15, and 18 qubits. In the sequel, the 6-vertex instance serves as the basic testing, whereas the 7- and 8-vertex instances function as verification of the method’s scalability. 

\subsection*{Basic testing}

We begin with the 6-vertex graph depicted in Fig. \ref{fig:6nodegraphpdom} to demonstrate the performance of the QAOA-based algorithm on PDP. The minimal DS of this graph is \( \{1, 4\} \), while the minimal PDS is either \( \{0, 4\} \) or \( \{1, 5\} \). Since this study focuses on solving the PDP rather than the DP, it was crucial to select a graph where the minimal DS and minimal PDS differ. 

\begin{figure}[htbp]
  \centering
  \begin{tikzpicture}[scale=1.2,
    every node/.style={circle, draw, minimum size=1em, inner sep=1pt},
    every edge/.style={draw, thick}]
    \node(0) at (1,1) {0};
    \node(1) at (1,2) {1};
    \node(2) at (2,2) {2};
    \node(3) at (1,3) {3};
    \node(4) at (2,3) {4};
    \node(5) at (3,3) {5};
    \draw (0) edge (1)
          (1) edge (2)
          (1) edge (3)
          (3) edge (4)
          (2) edge (4)
          (4) edge (5);
  \end{tikzpicture}
  \caption {A graph with 6 vertices and 6 edges}
  \label{fig:6nodegraphpdom}
\end{figure}

According to the conversion method outlined in Section \ref{sec:problemmodeling}, the QUBO model for the PDP of this graph is expressed in Eq. \ref{eq:expqubopdom}. Modeling this graph requires a total of 14 qubits. The corresponding Hamiltonian is derived by first substituting \( x_{*} \) with \( s_{*} \) as per Eq. \ref{eq:sx}, and then replacing \( s_{*} \) with \( \hat{\sigma}^{z} \). For simplicity, the detailed expansion of the Hamiltonian is omitted here.

\begin{equation}
  \begin{split}
    minimize \quad & x_0 + x_1 + x_2 + x_3 + x_4 + x_5 \\
    &+ P_{1} \cdot (1 - x_0 - x_1 + x_0 \cdot x_1) \\
    &+ P_{1} \cdot (x_1 + x_0 + x_2 + x_3 - (x_6 + 2 \cdot x_7) - 1)^2 \\
    &+ P_{1} \cdot (x_2 + x_1 + x_4 - (x_8 + x_9) - 1)^2 \\
    &+ P_{1} \cdot (x_3 + x_1 + x_4 - (x_{10} + x_{11}) - 1)^2 \\
    &+ P_{1} \cdot (x_4 + x_2 + x_3 + x_5 - (x_{12} + 2 \cdot x_{13}) - 1)^2 \\
    &+ P_{1} \cdot (1 - x_5 - x_4 + x_5 \cdot x_4) \\
    &+ P_{2} \cdot (x_0 \cdot (1 - x_1) + x_1 \cdot (1 - x_0) + x_1 \cdot (1 - x_2) \\ 
    & \quad + x_2 \cdot (1 - x_1) + x_1 \cdot (1 - x_3) 
    + x_3 \cdot (1 - x_1) \\
    & \quad + x_2 \cdot (1 - x_4) + x_4 \cdot (1 - x_2) + x_3 \cdot (1 - x_4) \\
    & \quad + x_4 \cdot (1 - x_3) + x_4 \cdot (1 - x_5) + x_5 \cdot (1 - x_4) - 6 \\
    & \quad + x_0 + x_1 + x_2 + x_3 + x_4 + x_5 ) \label{eq:expqubopdom}.
  \end{split}
  \end{equation}

\begin{figure}[htbp]
  \centering
  \includegraphics[height = 8cm, width=12.5cm]{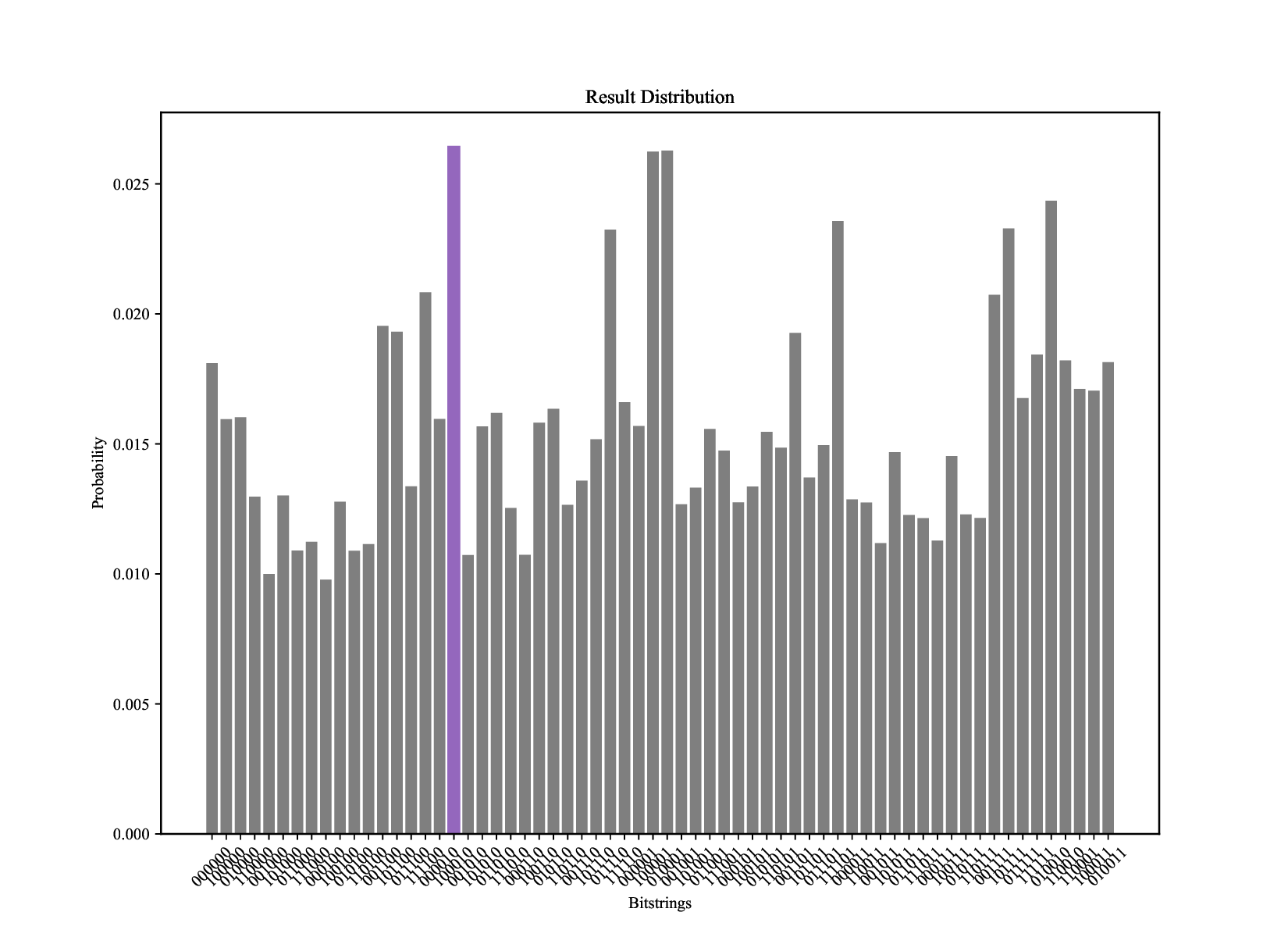}
  \caption {The probability distribution of bit strings when $q = 1$, $P_{1} = 7.2$, $P_{2} = 7.2$ and maximal iterations = 100}
  \label{fig:1-0.3-7.2-7.2-100}
\end{figure}

\begin{figure}[htbp]
  \centering
  \includegraphics[height = 8cm, width=12.5cm]{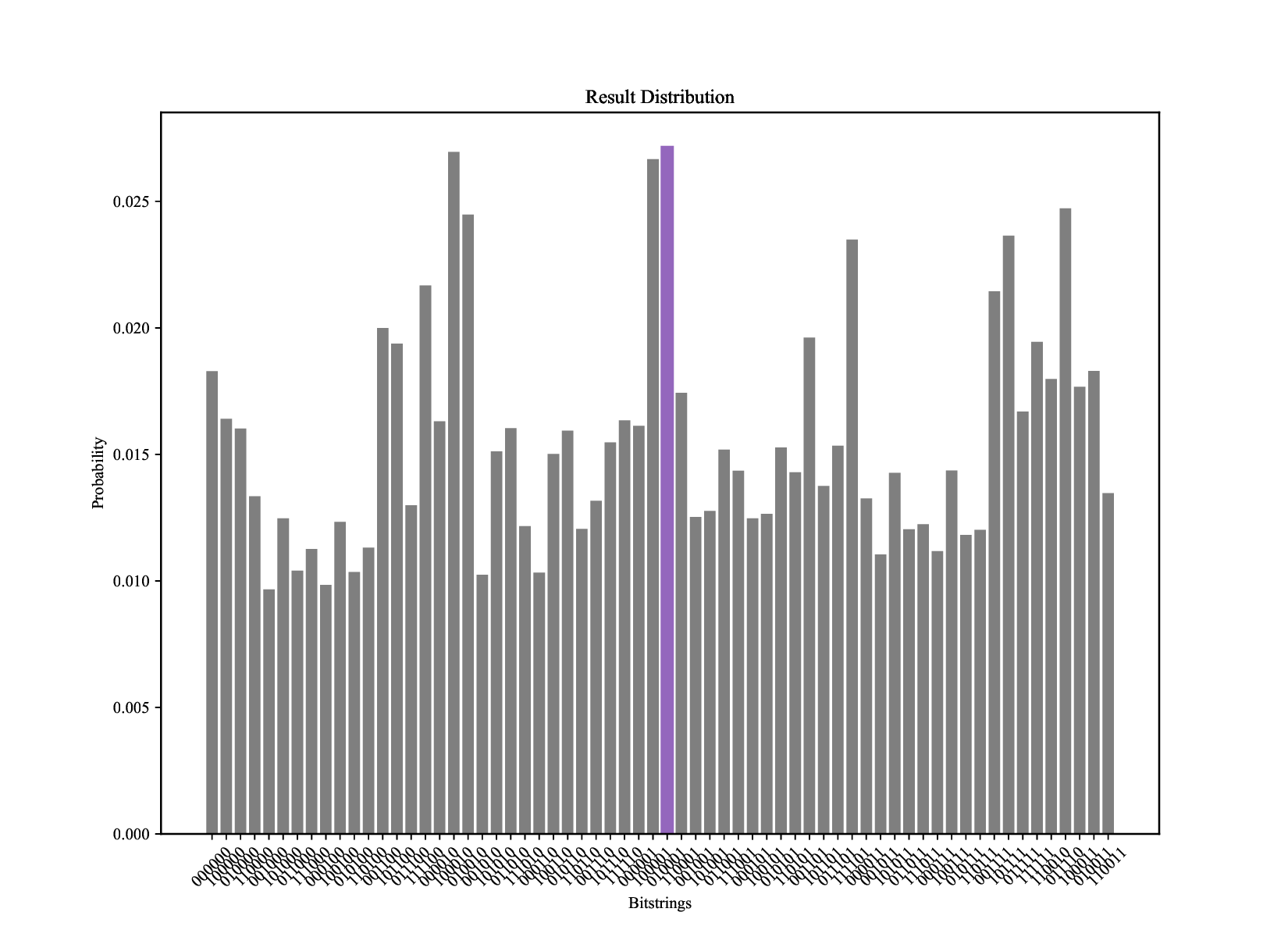}
  \caption {The probability distribution of bit strings when $q = 1$, $P_{1} = 7.2$, $P_{2} = 7.2$ and maximal iterations = 200}
  \label{fig:1-0.3-7.2-7.2-200}
\end{figure}

A total of 420 parameter combinations were tested, with \( q \in \{1, 2, 5\} \), \( P_{1} \in \{ 0.8, 1, 1.2, 1.4, 1.6, 1.8, 2 \} \times |V| \), \( \text{rate} = \frac{P_{2}}{P_{1}} \in \{0.3, 0.5, 0.7, 1\} \), and the maximum iterations set to \( \{100, 200, 500, 1000, 10000\} \). For the range of \( P_{1} \), the upper bound of the PDS was considered to be \( |V| \). Based on recommendations from \cite{RN416}, penalty coefficients were initially set between 0.7 and 1.5 times the value of the original objective function, with an extended range of \( [1.6, 2] \times |V| \). For \( P_{2} \), since the validity of constraint \ref{eq:pdpc2} depends on constraint \ref{eq:pdpc1}, we tested cases where \( P_{1} \leq P_{2} \). Out of the 420 parameter configurations examined, 82 yielded an output \( z_{*} \) satisfying the PDS condition, while 17 produced a \( z_{*} \) corresponding to the optimal PDS. Figures \ref{fig:1-0.3-7.2-7.2-100} and \ref{fig:1-0.3-7.2-7.2-200} illustrate the probability distributions of bit strings under the conditions \( q = 1 \), \( P_{1} = 7.2 \), \( P_{2} = 7.2 \), and maximal iterations of 100 and 200, respectively. In these figures, the probabilities of \( z_{*} \) are highlighted in purple. Notably, in both figures, \( z = 100010 \) and \( z = 010001 \) emerge as the two most probable bit strings, which correspond exactly to the two optimal PDS for the graph depicted in Fig. \ref{fig:6nodegraphpdom} (Figs. \ref{fig:d1opt}, \ref{fig:d2opt}). This alignment demonstrates that the final sampling results capture the symmetry inherent to the graph. Furthermore, increasing the number of iterations tends to reduce the probabilities of non-optimal bit strings, such as \( z = 100001 \) and \( z = 011110 \) in Fig. \ref{fig:1-0.3-7.2-7.2-100}. Although the performance of QAOA at low layers is limited, resulting in probabilities for \( z = 100010 \) and \( z = 010001 \) not being significantly higher than other bit strings, the experimental results suggest that with carefully chosen parameters, even low-layer QAOA can produce the expected results. These findings indicate that QAOA holds significant potential for solving the PDP.


\begin{figure}[htbp]
  \centering
  \begin{tikzpicture}[scale=1.2,
    every node/.style={circle, draw, minimum size=1em, inner sep=1pt},
    every edge/.style={draw, thick}]
    \node[red, fill=red!20](0) at (1,1) {0};
    \node(1) at (1,2) {1};
    \node(2) at (2,2) {2};
    \node(3) at (1,3) {3};
    \node[red, fill=red!20](4) at (2,3) {4};
    \node(5) at (3,3) {5};
    \draw (0) edge (1)
          (1) edge (2)
          (1) edge (3)
          (3) edge (4)
          (2) edge (4)
          (4) edge (5);
  \end{tikzpicture}
  \caption {Visualization of the $z = 100010$ with PDS $\{0,4\}$}
  \label{fig:d1opt}
\end{figure}

\begin{figure}[htbp]
  \centering
  \begin{tikzpicture}[scale=1.2,
    every node/.style={circle, draw, minimum size=1em, inner sep=1pt},
    every edge/.style={draw, thick}]
    \node(0) at (1,1) {0};
    \node[red, fill=red!20](1) at (1,2) {1};
    \node(2) at (2,2) {2};
    \node(3) at (1,3) {3};
    \node(4) at (2,3) {4};
    \node[red, fill=red!20](5) at (3,3) {5};
    \draw (0) edge (1)
          (1) edge (2)
          (1) edge (3)
          (3) edge (4)
          (2) edge (4)
          (4) edge (5);
  \end{tikzpicture}
  \caption {Visualization of the $z = 010001$ with PDS $\{1,5\}$.}
  \label{fig:d2opt}
\end{figure}

\begin{figure}[H]
  \centering
  \includegraphics[width=8cm]{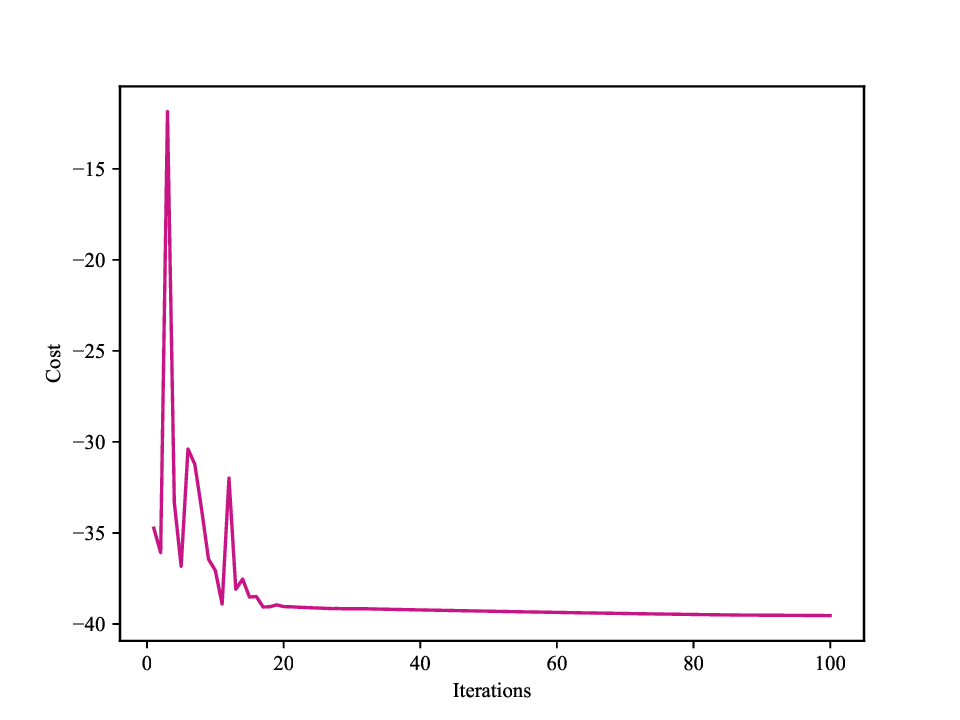}
  \caption {Cost of $q = 1$, $P_{1} = 7.2$, $P_{2} = 7.2$ and maximal iterations = 100}
  \label{fig:cost-1-0.3-7.2-7.2-100}
\end{figure}

\begin{figure}[H]
  \centering
  \includegraphics[width=8cm]{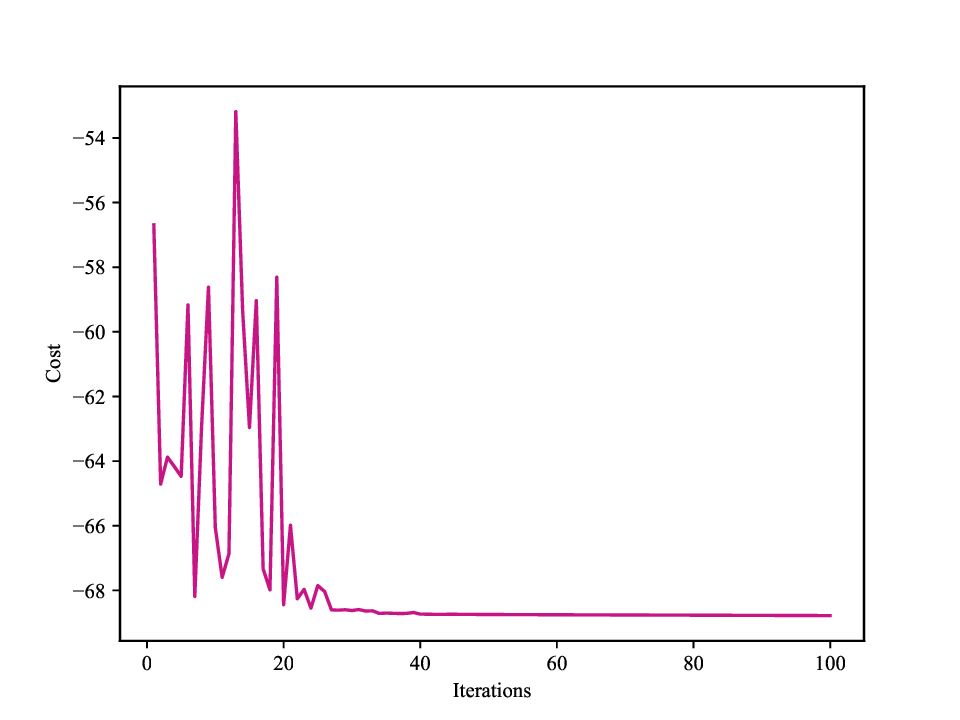}
  \caption {Cost of $q = 2$, $P_{1} = 12$, $P_{2} = 6$ and maximal iterations = 100}
  \label{fig:cost-2-0.3-12-6.0-100}
\end{figure}

\begin{figure}[H]
  \centering
  \includegraphics[width=8cm]{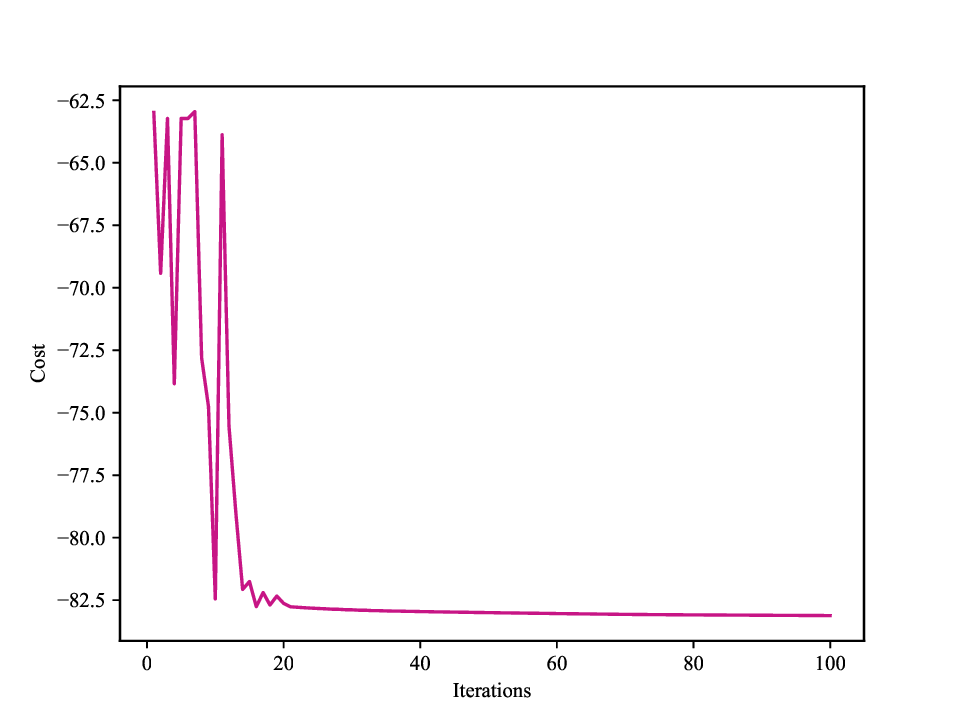}
  \caption {Cost of $q = 1$, $P_{1} = 12$, $P_{2} = 6$ and maximal iterations = 100}
  \label{fig:cost-1-0.3-12-6.0-100}
\end{figure}

To evaluate the convergence capability, we present the cost variation curves under different parameters in Figs. \ref{fig:cost-1-0.3-7.2-7.2-100}, \ref{fig:cost-2-0.3-12-6.0-100}, and \ref{fig:cost-1-0.3-12-6.0-100}. These figures reveal that the cost converges within approximately 20 iterations, highlighting a key advantage of employing low-layer QAOA to solve the PDP.

\begin{figure}[H]
  \centering
  \includegraphics[width=12cm]{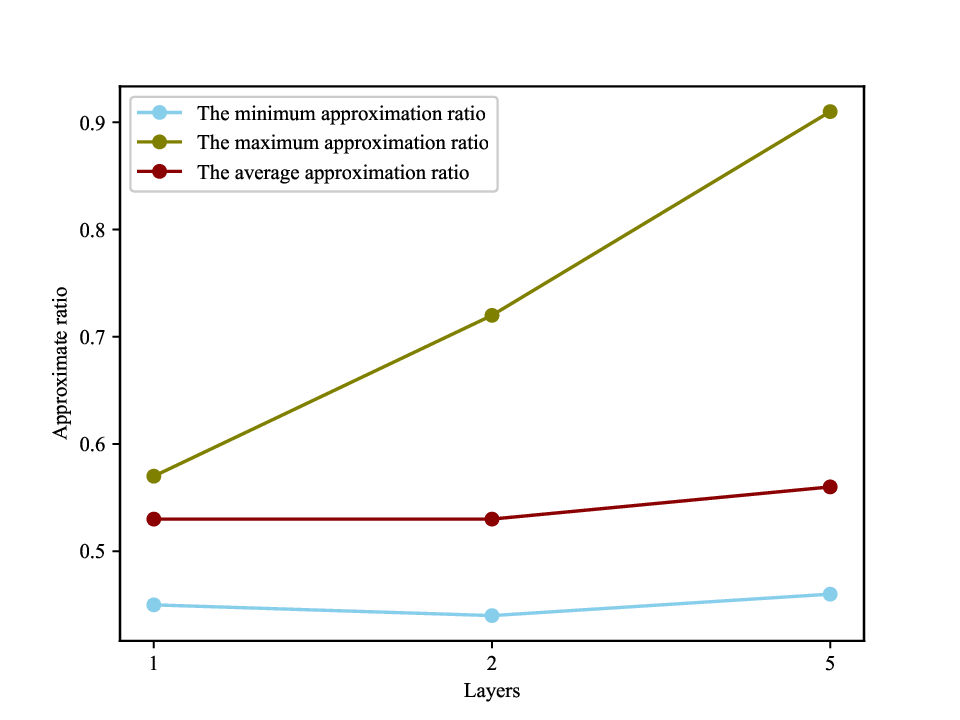}
  \caption {Minimal, maximal and average approximate ratios of different layers}
  \label{fig:arcompare}
\end{figure}

Next, we computed and analyzed the approximation ratios for the 420 experiments. Given that the PDP is a constrained minimization combinatorial optimization problem, bit strings that do not satisfy the constraints were excluded when calculating the approximation ratio \citep{RN502}. The approximation ratio formula used is presented in Eq. \ref{eq:ar}, where \( |PDS_{opt}| \) denotes the size of the optimal PDS, \( i \) is the index of the bit string satisfying the PDS condition, \( c_{i} \) represents the number of samples for the \( i \)-th bit string, \( |PDS_{i}| \) is the size of the PDS represented by the \( i \)-th bit string, and \( N_{total} \) is the total number of samples. In Fig. \ref{fig:arcompare}, we present the minimal, maximal, and average approximation ratios for different layer numbers. It is evident that as the number of layers increases, the approximation ratio exhibits an overall upward trend. When the number of layers reaches 5, the highest approximation ratio achieves approximately 0.9, clearly demonstrating the effectiveness of QAOA in approximating the solution to the PDS. Additionally, we observed that the upward trends for the minimal and average approximation ratios are less pronounced compared to the maximal approximation ratio. This disparity suggests that the maximal approximation ratio is more sensitive to the choice of parameters. Consequently, further analysis of the parameter dependency of the approximation ratio is crucial, as it would provide valuable insights for selecting optimal parameters to enhance the performance of QAOA in solving the PDP.

\begin{equation}
  R = \frac{|PDS_{opt}|}{ (\frac{\sum_{i} c_{i} \cdot |PDS|_{i}}{N_{total}})} \label{eq:ar}.
  \end{equation}

\begin{figure}[htbp]
    \centering
    \begin{subfigure}[b]{0.48\textwidth}
        \centering
        \begin{tikzpicture}
          \draw pic[scale=2] {piechart2={-126}{
              26.2/ 1              /myblue,
              32.1/ 2           /myred,
              41.7/ 5       /mygreen%
          }};
        \end{tikzpicture}
        \caption{Layer}
    \end{subfigure}
    \hfill 
    \begin{subfigure}[b]{0.48\textwidth}
        \centering
        \begin{tikzpicture}
          \draw pic[scale=2] {piechart2={-90}{
              22.6/ 100                /myblue,
              21.4/ 200                /myred,
              17.9/ 500                /mygreen,
              17.9/ 1000             /mydarkpurple,
              20.2/ 10000       /myyellow%
          }};
        \end{tikzpicture}
        \caption{Max iterations}
    \end{subfigure}
    \begin{subfigure}[b]{0.48\textwidth}
        \centering
        \begin{tikzpicture}
          \draw pic[scale=2] {piechart2={-90}{
              27.4/ 0.3                /myblue,
              22.6/ 0.5                /myred,
              15.5/ 0.7                /mygreen,
              34.5/ 1             /mydarkpurple%
          }};
        \end{tikzpicture}
        \caption{$\frac{P_{2}}{P_{1}}$}
    \end{subfigure}
    \hfill 
    \begin{subfigure}[b]{0.48\textwidth}
        \centering
        \begin{tikzpicture}
          \draw pic[scale=2] {piechart2={-90}{
              7.1/ 4.8                /myblue,
              11.9/ 6.0                /myred,
              29.8/ 7.2                /mygreen,
              11.9/ 8.4             /mydarkpurple,
              8.3/ 9.6       /myyellow,
              26.2/ 10.8             /mypink,
              4.8/ 12.0       /mydarkestblue%
          }};
        \end{tikzpicture}
        \caption{$P_{1}$}
    \end{subfigure}
    \caption{Analysis of the parameter distribution in the top 20\% based on approximate ratio}
    \label{fig:arpie}
\end{figure}

We sorted the approximation ratios for the 420 experiments in descending order and selected the top 20\% (84 experiments) with the highest approximation ratios for further analysis. For each parameter set, we counted the frequency of its occurrence among these 84 experiments. The statistical results are presented in Fig. \ref{fig:arpie}. It is evident that experiments achieving higher approximation ratios are strongly correlated with higher layer numbers, while no clear trend is observed concerning the maximum number of iterations. Regarding the weight parameters, setting \( P_{1} \) to 1.8 times \( |V| \) and \( P_{2} \) equal to \( P_{1} \) appears to be more favorable for attaining higher approximation ratios. The observation that higher approximation ratios require higher values of \( P_{1} \) and \( P_{2} \) is reasonable. In the model, emphasizing the satisfaction of the DS condition (\( P_{1} \)) and the perfect condition (\( P_{2} \)) aligns with the pursuit of higher approximation ratios, as a PDS is, by definition, a DS that satisfies the perfect condition. However, when calculating the approximation ratio, erroneous bit strings must be excluded, so we cannot assume that experiments with the highest approximation ratios necessarily yield a higher probability for the optimal PDS compared to all other bit strings (including non-PDS bit strings). Instead, we can infer that the probability of the optimal PDS has a relative advantage locally, compared to other non-optimal PDS probabilities. This concern stems from two main factors: (1) excessively high penalty weights can prevent the original objective function from converging to a desirable solution \citep{RN416}, and (2) constraint \ref{eq:pdpc2} involves overlapping qubits with all vertices in constraint \ref{eq:pdpc1}, leading to mutual influence, particularly when the two constraints are assigned similar weights. Subsequent experimental results will further validate this analysis.

In Figs. \ref{fig:optpie} and \ref{fig:corpie}, we present the statistical distribution of parameters for all experiments where \( z_{*} \) corresponds to either an optimal PDS or a correct PDS. The statistical data indicate that smaller layer numbers are more conducive to allowing the optimal PDS and correct PDS to dominate the overall probability distribution. Similar to the approximation ratio, these results appear to be independent of the maximum number of iterations. Regarding the weights, setting \( P_{1} \) to 2 times the number of vertices and \( \frac{P_{2}}{P_{1}} = 0.5 \) provides a notable numerical advantage. In comparison to the approximation ratio, these results differ in terms of layer numbers and the ratio of \( P_{2} \) to \( P_{1} \). Larger layer numbers may pose a challenge for the optimization algorithm, as the number of angles it must adjust increases proportionally, being twice the number of layers. Consequently, higher layer numbers and larger penalty values are more likely to lead the algorithm into a local optimum. To improve the overall probability of obtaining the optimal PDS, selecting discriminative and well-tuned values for \( P_{1} \) and \( P_{2} \) might be a more effective approach.


\begin{figure}[htbp]
    \centering
    \begin{subfigure}[b]{0.48\textwidth}
        \centering
        \begin{tikzpicture}
          \draw pic[scale=2] {piechart2={-126}{
              70.6/ 1              /myblue,
              11.8/ 2           /myred,
              17.6/ 5       /mygreen%
          }};
        \end{tikzpicture}
        \caption{Layer}
    \end{subfigure}
    \hfill 
    \begin{subfigure}[b]{0.48\textwidth}
        \centering
        \begin{tikzpicture}
          \draw pic[scale=2] {piechart2={-90}{
              17.6/ 100                /myblue,
              23.5/ 200                /myred,
              23.5/ 500                /mygreen,
              23.5/ 1000             /mydarkpurple,
              11.8/ 10000       /myyellow%
          }};
        \end{tikzpicture}
        \caption{Max iterations}
    \end{subfigure}
    \begin{subfigure}[b]{0.48\textwidth}
        \centering
        \begin{tikzpicture}
          \draw pic[scale=2] {piechart2={-90}{
              17.6/ 0.3                /myblue,
              52.9/ 0.5                /myred,
              29.4/ 1             /mydarkpurple%
          }};
        \end{tikzpicture}
        \caption{$\frac{P_{2}}{P_{1}}$}
    \end{subfigure}
    \hfill 
    \begin{subfigure}[b]{0.48\textwidth}
        \centering
        \begin{tikzpicture}
          \draw pic[scale=2] {piechart2={-90}{
              17.6/ 6.0                /myred,
              29.4/ 7.2                /mygreen,
              11.8/ 10.8             /mypink,
              41.2/ 12.0       /mydarkestblue%
          }};
        \end{tikzpicture}
        \caption{$P_{1}$}
    \end{subfigure}
    \caption{Analysis of the parameter distribution of the experiments of which $z^{*}$ is optimal (Due to rounding, the sum of probabilities for some pie charts is 99.9\%)}
    \label{fig:optpie}
\end{figure}

\begin{figure}[htbp]
    \centering
    \begin{subfigure}[b]{0.48\textwidth}
        \centering
        \begin{tikzpicture}
          \draw pic[scale=2] {piechart2={-126}{
              39.3/ 1              /myblue,
              26.2/ 2           /myred,
              34.5/ 5       /mygreen%
          }};
        \end{tikzpicture}
        \caption{Layer}
    \end{subfigure}
    \hfill 
    \begin{subfigure}[b]{0.48\textwidth}
        \centering
        \begin{tikzpicture}
          \draw pic[scale=2] {piechart2={-90}{
              22.6/ 100                /myblue,
              22.6/ 200                /myred,
              22.6/ 500                /mygreen,
              19.0/ 1000             /mydarkpurple,
              13.1/ 10000       /myyellow%
          }};
        \end{tikzpicture}
        \caption{Max iterations}
    \end{subfigure}
    \begin{subfigure}[b]{0.48\textwidth}
        \centering
        \begin{tikzpicture}
          \draw pic[scale=2] {piechart2={-90}{
              17.9/ 0.3                /myblue,
              35.7/ 0.5                /myred,
              11.9/ 0.7                /mygreen,
              34.5/ 1             /mydarkpurple%
          }};
        \end{tikzpicture}
        \caption{$\frac{P_{2}}{P_{1}}$}
    \end{subfigure}
    \hfill 
    \begin{subfigure}[b]{0.48\textwidth}
        \centering
        \begin{tikzpicture}
          \draw pic[scale=2] {piechart2={-90}{
              13.1/ 4.8                /myblue,
              13.1/ 6.0                /myred,
              9.5/ 7.2                /mygreen,
              20.2/ 8.4             /mydarkpurple,
              16.7/ 9.6       /myyellow,
              9.5/ 10.8             /mypink,
              17.9/ 12.0       /mydarkestblue%
          }};
        \end{tikzpicture}
        \caption{$P_{1}$}
    \end{subfigure}
    \caption{Analysis of the parameter distribution of the experiments of which  $z^{*}$ is correct (Due to rounding, the sum of probabilities for some pie charts is 99.9\%)}
     \label{fig:corpie}
\end{figure}


Based on the analysis of Figs. \ref{fig:arpie}, \ref{fig:optpie}, and \ref{fig:corpie}, we conclude that when using QAOA to solve the PDP, pursuing a higher approximation ratio and increasing the probability of sampling the optimal PDS can be two distinct objectives. To enable QAOA to directly output \( z_{*} \) as a PDS, particularly the optimal PDS, it is advisable to use a smaller number of layers and assign distinct weights to the two types of penalties. Conversely, if an additional filtering step can be applied to the final sampling results, a larger number of layers can be used, with both penalty weights set to equal and higher values. This approach would be more effective in achieving a higher approximation ratio.

\begin{figure}[htbp]
  \centering
  \includegraphics[width=14cm]{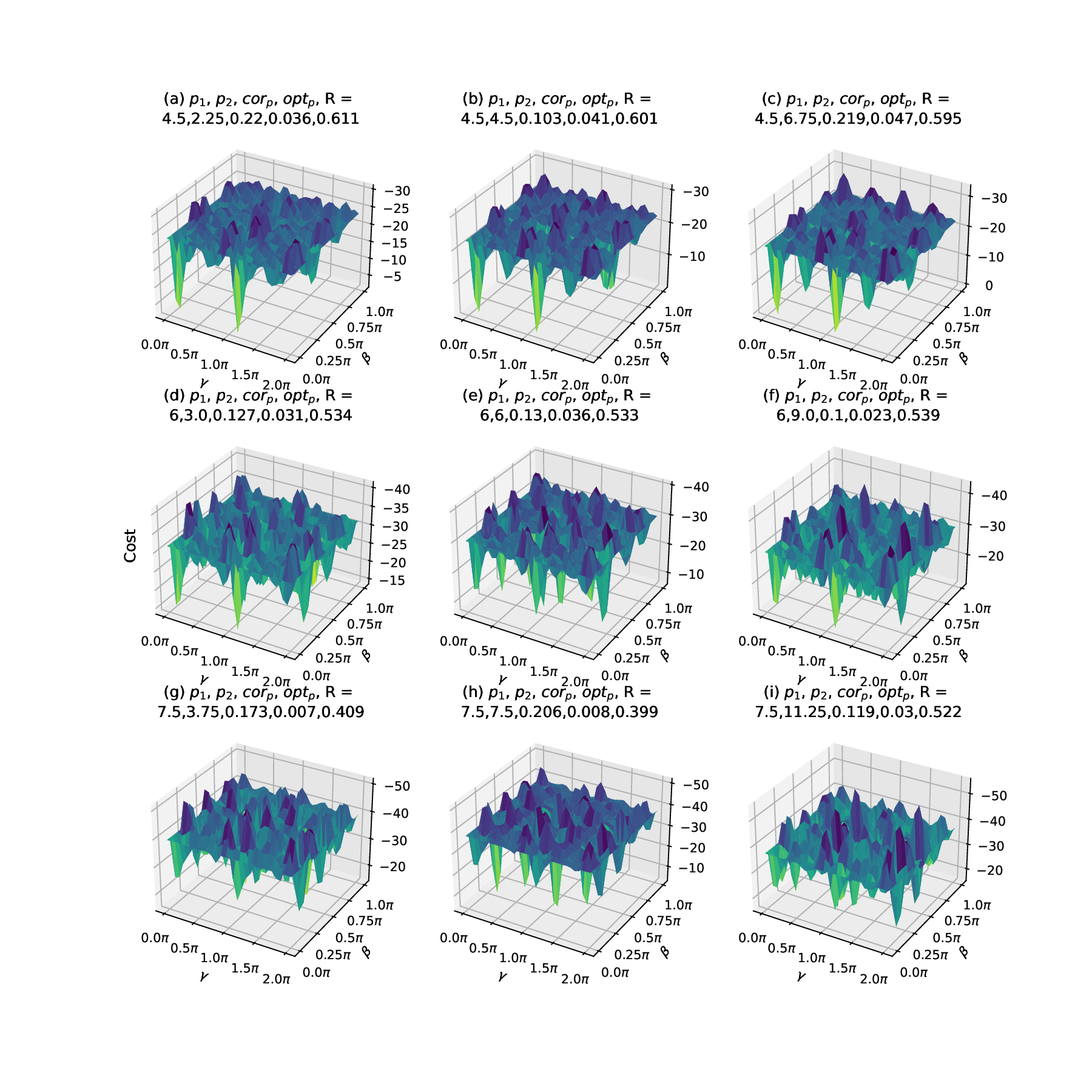}
  \caption {Landscape maps of single layer QAOA within different $P_{1}$ and $P_{2}$}
  \label{fig:land}
\end{figure}

As a supplement to the case where \( P_2 > P_1 \), in Fig. \ref{fig:land}, we present the landscape maps of the single layer QAOA under the conditions where \( \frac{P_2}{P_1} \in \{0.5, 1, 1.5\} \) and \( P_1 \in \{0.75, 1, 1.25\} \times |V| \). For the convenience of observation, the vertical axis has been flipped. Here, \( \gamma \in [0, 2\pi] \) and \( \beta \in [0, \pi] \), while \( cor_p \) and \( opt_p \) represent the correct probability and optimal probability, respectively. Although we cannot draw further conclusions regarding the value of \( P_1 \), based on the data in Fig. \ref{fig:land}, we can derive three interesting observations. First, the correct probabilities and optimal probabilities reflected in subfigures (a), (b), and (c) are superior to those of the other six parameter combinations. In terms of the optimization landscape, the number of local optimal solutions in these three subfigures is generally fewer than that in the other six subfigures, and the overall distribution of the optimization landscape is smoother. This may be the reason why their optimization results are better than those of other parameter combinations. Second, \( P_1 \) has a greater impact on the global probability than \( P_2 \), as the optimization landscapes of the three subfigures in the same row are similar, and they share the same \( P_1 \). Third, by observing the data in the second row (d, e, f) and the third row (g, h, i), it can be found that even when the global correct probability and optimal probability are not high, setting a larger \( P_2 \) is beneficial for achieving a better approximation ratio, which also confirms our previous analysis.

\subsection*{Additional cases}

\begin{figure}[htbp]
  \centering
  \includegraphics[width=9cm]{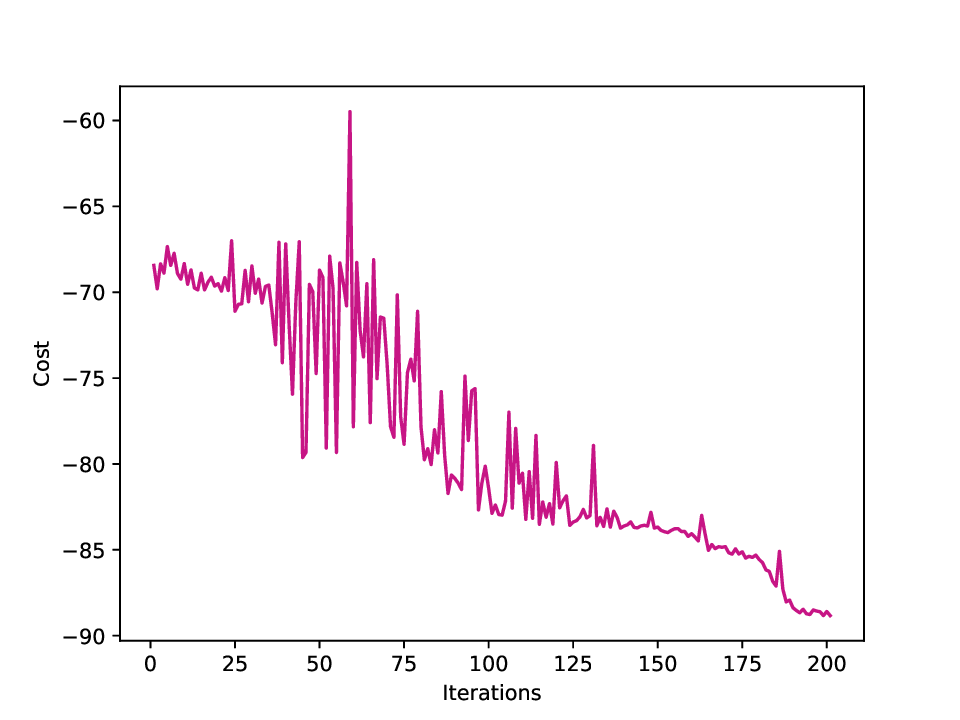}
  \caption { Cost of the graph with 7 vertices when $q = 10$, $P_{1} = 10.5$, $P_{2} = 5.25$ and maximal iterations = 200}
  \label{fig:7cost}
\end{figure}

\begin{figure}[htbp]
  \centering
  \includegraphics[width=9cm]{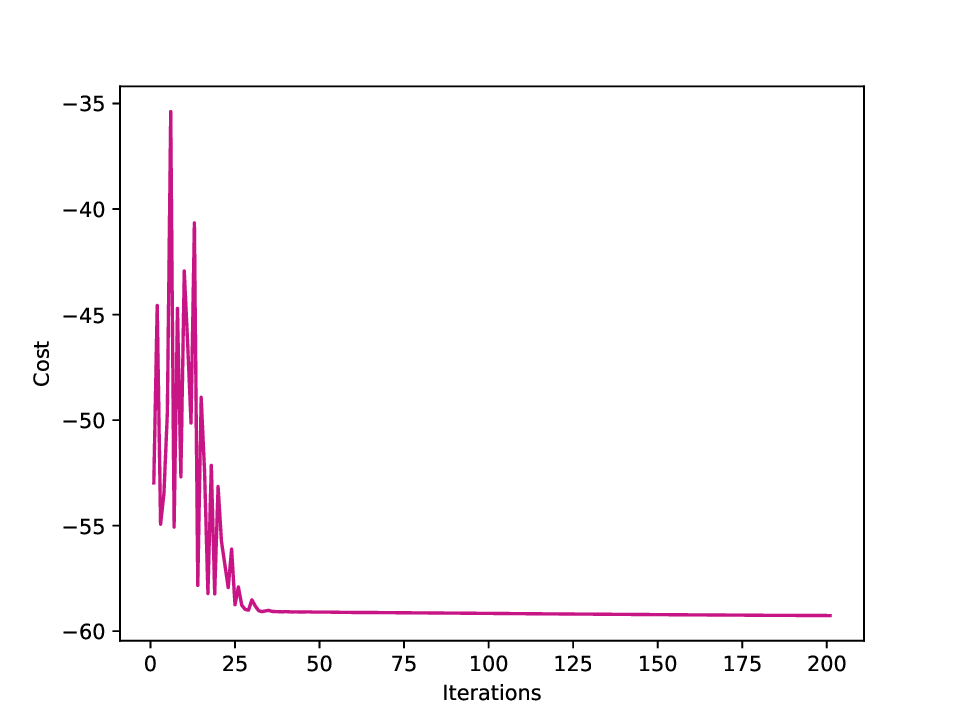}
  \caption {Cost of the graph with 8 vertices when $q = 2$, $P_{1} = 8$, $P_{2} = 4$ and maximal iterations = 200}
  \label{fig:8cost}
\end{figure}

\begin{figure}[htp]
  \centering
  \includegraphics[width=11.4cm]{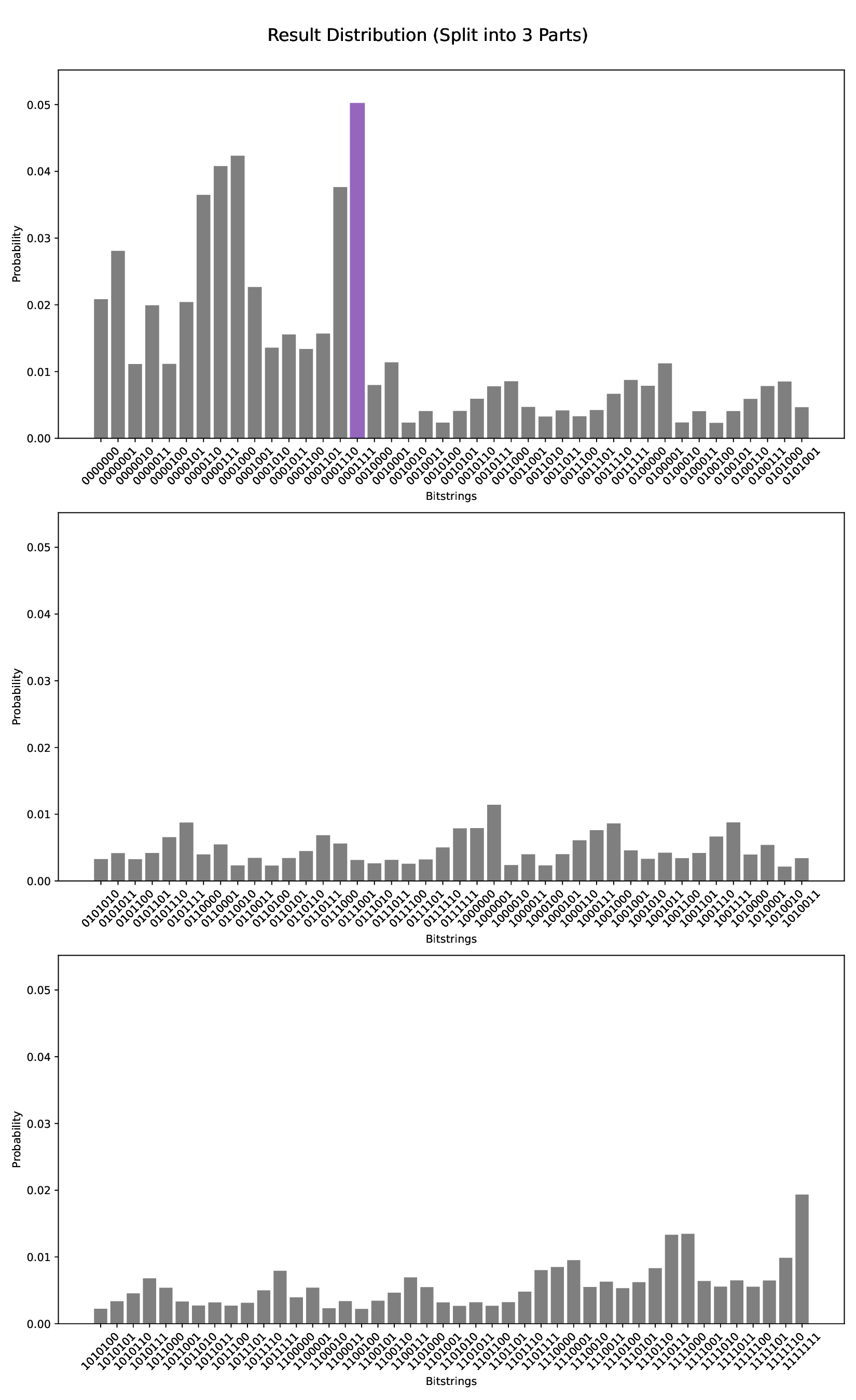}
  \caption {Sampling probability distribution for the 7-vertex graph obtained with \(q = 10\), \(P_{1} = 10.5\), \(P_{2} = 5.25\), and a maximum of 200 iterations}
  \label{fig:7d}
\end{figure}

\begin{figure}[htp]
  \centering
  \includegraphics[width=11.4cm]{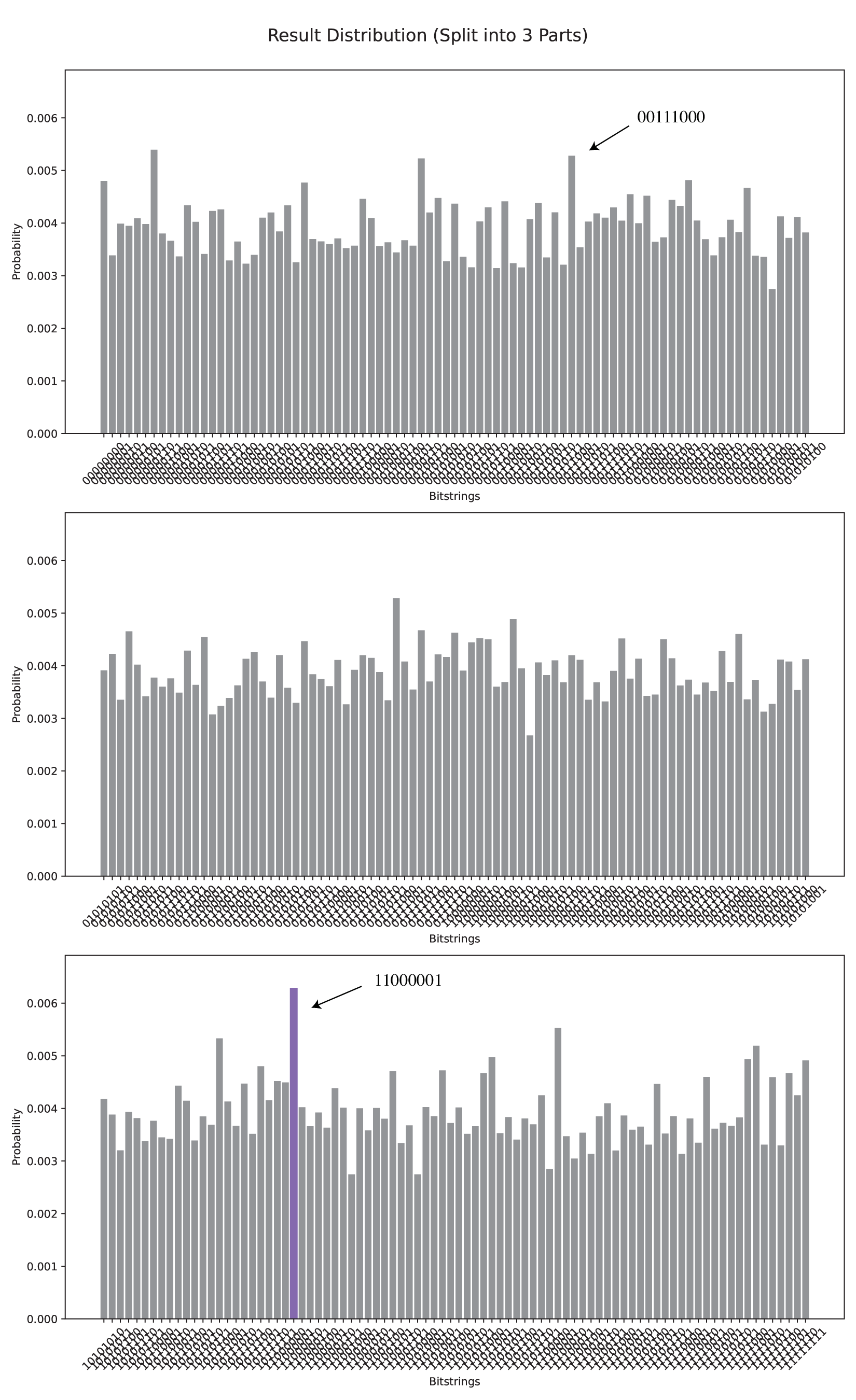}
   \caption {Sampling probability distribution for the 8-vertex graph obtained with \(q = 2\), \(P_{1} = 8\), \(P_{2} = 4\), and a maximum of 200 iterations}
  \label{fig:8d}
\end{figure}

We further evaluated the proposed approach on larger graphs, namely those depicted in Fig. \ref{fig:7node_graph} (7 vertices) and Fig. \ref{fig:8node_graph} (8 vertices), requiring 15 and 18 qubits, respectively.  
For the 7-vertex instance, the optimal PDS is \(\{3,4,5,6\}\), corresponding to the bitstring \(z^{*} = 0001111\); for the 8-vertex instance, the optimal PDSs are \(\{0,1,7\}\) and \(\{2,3,4\}\), yielding either \(z^{*} = 11000001\) or \(00111000\). The final sampling distributions for the two graphs are presented in Fig. \ref{fig:7d} and Fig. \ref{fig:8d}, and the associated cost trajectories are shown in Fig. \ref{fig:7cost} and Fig. \ref{fig:8cost}, respectively.

With respect to the sampling distributions, the bit strings exhibiting the highest empirical probabilities (highlighted in purple) coincide with the global optima in both test cases. For the 7-vertex instance (Fig. \ref{fig:7d}), the optimal bit string \(z^{*} = 0001111\) commands a pronounced global advantage in probability mass. In the 8-vertex case (Fig. \ref{fig:8d}), the optimal string \(z^{*} = 11000001\) likewise attains a clear supremacy, whereas the alternative optimum \(\{2,3,4\}\) encoded as \(00111000\) is not the second most frequently sampled configuration; nevertheless, a discernible local peak in its vicinity is observed.

Comparatively, the peak sampling probabilities for the 8-vertex graph are approximately one order of magnitude lower than those for the 7-vertex graph. This degradation may be attributed to two concurrent factors: the enlarged graph size compromises optimization efficacy, and the increased qubit count expands the Hilbert space, thereby diluting the probability density of any single computational basis state.

Regarding the cost trajectories, although Fig. \ref{fig:8cost} exhibits ostensibly superior convergence behaviour relative to Fig. \ref{fig:7cost}, the optimizer evidently stalls at a sub-optimal local minimum, as evidenced by the failure of the second optimum \(00111000\) to rank among the two most probable outcomes. Conversely, the trend in Fig. \ref{fig:7cost} suggests that further iterations would likely continue to amplify the probability mass on \(0001111\).

Overall, within the investigated regime graphs of up to eight vertices and quantum circuits of up to eighteen qubits, the results remain consistent with theoretical expectations.

\begin{figure}[htbp]
  \centering
  \begin{tikzpicture}[scale=1.2,
    every node/.style={circle, draw, minimum size=1em, inner sep=1pt},
    every edge/.style={draw, thick}]
    \node (0) at (1,1) {0};
    \node (1) at (2,1) {1};
    \node (2) at (3,1) {2};
    \node (3) at (2,2) {3};
    \node (4) at (1.5,3) {4};
    \node (5) at (2.5,3) {5};
    \node (6) at (2,4) {6};
    \draw (0) edge (3)
          (1) edge (3)
          (2) edge (3)
          (3) edge (4)
          (3) edge (5)
          (4) edge (5)
          (5) edge (6)
          (4) edge (6);
  \end{tikzpicture}
  \caption{A graph with 7 vertices and 8 edges}
  \label{fig:7node_graph}
\end{figure}

\begin{figure}[htbp]
  \centering
  \begin{tikzpicture}[scale=1.2,
    every node/.style={circle, draw, minimum size=1em, inner sep=1pt},
    every edge/.style={draw, thick}]
    \node (0) at (1,1)   {0};
    \node (1) at (4,1)   {1};
    \node (2) at (2.5,5) {2};
    \node (3) at (2,2)   {3};
    \node (4) at (3,2)   {4};
    \node (5) at (3,3)   {5};
    \node (6) at (2,3)   {6};
    \node (7) at (2.5,4) {7};
    \draw (0) edge (3)
          (3) edge (4)
          (4) edge (5)
          (5) edge (7)
          (6) edge (7)
          (3) edge (6)
          (2) edge (7)
          (1) edge (4);
  \end{tikzpicture}
  \caption{A graph with 8 vertices and 8 edges}
  \label{fig:8node_graph}
\end{figure}

\section{Conclusion}\label{sec:conclusion}

In this paper, we explored the use of QAOA to solve the PDP. We began by modeling the PDP as a 0-1 integer programming problem based on its definition and transformed its two types of constraints into quadratic penalties to derive the QUBO model. Through variable substitution, we ultimately obtained the Hamiltonian for the PDP. Using IBM's Qiskit quantum simulator, we conducted extensive tests and analyses on the performance of QAOA in solving the PDP. 

In basic testing, we tested 420 parameter combinations involving layer numbers, penalty coefficients, and maximum iterations. On a macro scale, QAOA successfully computed the correct PDS in 82 parameter combinations and identified the optimal PDS in 17 combinations, achieving a highest approximation ratio of approximately 0.9. These results confidently demonstrate that even with low-layer QAOA, excellent outcomes can be achieved with appropriately chosen parameters. This underscores the significant potential of quantum algorithms in solving the PDP. In the parameter analysis, we primarily examined the parameter distributions for three categories of experiments: (1) those in the top 20\% for approximation ratio, (2) those where \( z_{*} \) corresponds to the optimal PDS, and (3) those where \( z_{*} \) corresponds to the correct PDS. The findings revealed distinct parameter tendencies between the first category and the latter two. Pursuing a high approximation ratio focuses on the advantage of relative probability, in which case there is a tendency to increase the penalty coefficient for the perfect condition. In contrast, the other two metrics reflect the advantage of global probability, where distinct penalty coefficient settings are more preferable, and QAOA with a small number of layers is better suited for traditional optimization algorithms to optimize the global probability under the current test cases. Subsequent analysis results on the differences in optimization landscapes under different ratios of penalty coefficients have also confirmed this conclusion. In addition to the 6-vertex graph, we also tested slightly larger graphs with 7 and 8 vertices. It can be observed that under the given parameters, the QAOA-based method can compute the optimal PDS in all cases. Although the order of magnitude of the probability of obtaining the optimal solution decreases, the advantage of the sampling probability of the optimal solution in the global context is quite evident. These results confirm the effectiveness of this method.These insights provide valuable guidance for optimizing parameter settings when applying QAOA to solve the PDP, paving the way for further advancements in quantum algorithm applications.

The limitations of this paper include:  
(1) Due to the substantial computational resources required for local quantum simulations, and given that modeling the PDP for an 8-vertex graph already approaches 18 qubits, this paper is currently unable to test larger-scale instances.  
(2) All experiments were performed exclusively on a noise-free classical simulator; neither execution on a physical quantum processor nor circuit-level noise-mitigation strategies were pursued.

Based on the research content, experimental results, and limitations discussed in this paper, we propose the following directions for future work: (1) Evaluate the effectiveness of QAOA in solving the PDP on a real quantum computer. (2) Incorporate the noise environment into the QAOA algorithm and optimize the quantum circuit accordingly. (4) Explore modeling approaches for the PDP that require fewer qubits. (5) Extend the application of QAOA to address other variants of the DP, such as the \( k \)-domination problem and related challenges.

\section*{Acknowledgement}

The experimental setup was inspired by the open-source project named qopt-best-practices maintained by the Qiskit community. We also gratefully acknowledge the technical support from China Communications Information \& Technology Group Co., Ltd.  

\section*{Data Availability}
The data used to support the findings of this study are included within the article.

\section*{Disclosure of interest}
The authors declare that they have no conflicts of interest that could have appeared to influence the work reported in this paper.

\section*{Funding statement}
 
This work is supported by Fundamental Research Funds for the Central Universities and National Natural Science Foundation of China (No. 12331014).

\section*{Declaration of Generative AI}

During the preparation of this work the authors used ChatGPT, DouBao and Kimi in order to improve readability and language. After using this tool, the authors reviewed and edited the content as needed and take full responsibility for the content of the publication.



 \bibliographystyle{elsarticle-harv} 
 \bibliography{pdom}






\end{CJK}
\end{document}